\documentclass[11pt]{article}
\usepackage{amsmath,amsfonts}
\usepackage{graphicx}   
\usepackage[margin=1in]{geometry}
\usepackage{color}  

\newcommand{\Cx}{{\mathbb C}}

\renewcommand{\H}{{\mathcal H}}

\def\idty{{\mathchoice {\mathrm{1\mskip-4mu l}} {\mathrm{1\mskip-4mu l}} %
{\mathrm{1\mskip-4.5mu l}} {\mathrm{1\mskip-5mu l}}}}

\newcommand{\be}{\begin{equation}}
\newcommand{\ee}{\end{equation}}
\newcommand{\bea}{\begin{eqnarray}}
\newcommand{\eea}{\end{eqnarray}}
\newcommand{\beann}{\begin{eqnarray*}}
\newcommand{\eeann}{\end{eqnarray*}}

\newtheorem{theorem}{Theorem}[section]

\newtheorem{lemma}[theorem]{Lemma}

\begin{document}

\title{Many-Body Localization: 
Concepts and Simple Models}

\author{Robert Sims\\[10pt]
Department of Mathematics\\ University of Arizona\\
Tucson, AZ 85721, USA\\
Email: rsims@math.arizona.edu \\[10pt]
and \\[10pt]
G\"unter Stolz\\[10pt]
Department of Mathematics\\ University of Alabama at Birmingham\\
Birmingham, AL 35294, USA\\
Email: stolz@uab.edu}

\bigskip

\maketitle

\abstract{We review recent results on many-body localization for two explicitly analyzable models of many-body quantum systems, the XY spin chain in transversal magnetic field as well as interacting systems of harmonic quantum oscillators. In both models the presence of disorder leads to dynamical localization in the form of zero-velocity Lieb-Robinson bounds and to exponential decay of ground state correlations. Moreover, for oscillator systems one can also show exponential decay of thermal states as well as an area law bound for the entanglement entropy of ground and thermal states. The key fact which allows a rigorous analysis of these models is that they are given by many-body Hamiltonians which can be reduced to effective single particle Hamiltonians.
}

\footnotetext[1]{Copyright \copyright\ 2013 by the authors. 
This paper may be reproduced, in its entirety, for non-commercial purposes.}

\section{What is many-body localization?}

While the theory of localization for the Anderson model and other types of {\it single-particle} random Schr\"odinger operators has been rigorously studied since the late 1970s, mathematical physicists have started to consider localization phenomena in interacting disordered {\it multi-particle} and {\it many-particle} quantum systems only much more recently. 

The most obvious reason for this is that understanding the combined effects of disorder and interactions on large quantum systems is difficult. In fact, the aim of arriving at a better understanding of the many-body localization transition, starting from its correct meaning, has been a very active topic in the recent theoretical physics literature (e.g.\ \cite{Gornyietal, Baskoetal, OganesyanHuse, Znidaricetal, PalHuse, Canovietal, VoskAltman}) and, even on the level of physical heuristics, is not yet fully understood. A nice introductory account of a physicist's view of many-body localization as well as of thermalization (the many-body analogue of extended states in single-particle systems) can be found in \cite{Huse2009}.

Rigorous work on the many-body localization transition has to start on the side of localization, given that extended states are not understood rigorously even in the one-particle setting. 

In the last half-decade several groups of authors have proven localization properties for the multi-particle Anderson model. This covers results for the discrete Anderson model \cite{AizenmanWarzel, ChulaevskySuhov2009a, ChulaevskySuhov2009b, KleinNguyen2012} as well as for the continuous Anderson model \cite{Chulaevskyetal2011, KleinNguyen2013}. In all of these works, a fixed (although arbitrary) particle number $N$ is considered and spectral as well as dynamical localization is proven for some of the physically expected regimes. No particular attempt is made to gain optimal control of appearing constants on $N$, nor would the methods used allow to do this. In other words, the authors consider what physicists refer to as {\it multi-particle systems}.

This should be distinguished from {\it many-body systems}. The latter ideally refers to systems of infinitely many interacting particles. Alternatively, one can consider systems of finitely many particles, prove properties for them which hold uniformly in the particle number $N$, which in turn might lead to results for the infinite system by letting $N$ go to infinity. Generally, in the quantum setting, the direct description of infinite particle systems requires concepts and techniques from operator algebras. In this introductory survey paper we will avoid going into these issues and thus focus entirely on localization bounds for finite systems, but with constants which are uniform in the particle number.

More specifically, we will discuss results of three types here which can be considered as many-body localization properties:

\begin{itemize}

\item Zero-velocity Lieb-Robinson bounds,

\item Exponential decay of correlations for ground states and thermal states,

\item Slow growth of the entanglement entropy (``area laws'') for ground states and thermal states. 

\end{itemize}

Zero-velocity Lieb-Robinson bounds are a form of many-body {\it dynamical} localization. Here it is important to stress from the outset that the latter does not refer to the absence of particle transport, but to the {\it absence of information transport} within a system of particles which essentially stay stationary. In this context the spreading or non-spreading of waves refers to group waves within the particle system.

In single particle theory, localization properties fall into two types, dynamical localization and spectral localization. In its most familiar form of dense pure point spectrum the latter can not be discussed in our setting of finite particle systems, at least not for the examples considered here, where all Hamiltonians will have purely discrete spectrum (not unlike the situation which occurs when studying one-particle random operators in large but finite volume). Instead of spectral properties we will thus concentrate on properties of eigenfunctions of disordered many-body systems. 

We think of exponential decay of correlations and of area laws as forms of eigenfunction localization, as they can be interpreted as saying that eigenstates of an interacting disordered many-body system are, in suitable sense, close to those of a non-interacting system (where eigenstates are simple product states, so that correlations as well as entanglement trivially vanish).

We need to be a bit more precise here: Exponential decay of ground state correlations and area laws have been proven for some classes of deterministic {\it uniformly gapped} many-body systems. This means that the gap size $E_1-E_0$ between the ground state energy and first excited energy of the many-body Hamiltonian is bounded from below by some $\gamma>0$, uniformly in the particle number. This does not require the presence of disorder and thus, strictly speaking, should not be considered as a localization phenomenon (at least not as Anderson localization). What is of interest to us here are situations where similar results can be proven without the assumption of a uniform spectral gap, but instead are due to a disorder induced {\it mobility gap}, see, e.g., \cite{Hastings2010} for a physical perspective.

Our main goal here is to survey the results of the papers \cite{HSS}, \cite{NSS1} and \cite{NSS2}, where two simple many-body models are studied which allow to prove the localization properties discussed above. These models are ``simple'' in the sense that their many-body localization properties can be reduced to the localization properties of effective single particle Hamiltonians. We consider the study of such models as a first step towards rigorously understanding many-body localization phenomena, which should be followed by the investigation of more realistic and mathematically much more difficult ``true'' many-body systems. This will pose the considerable challenge of finding new methods to prove localization which work directly in the many-body context. 

A promising class of models to understand the disorder effects on many-body systems are quantum spin systems. They are considered in many of the physics references mentioned above and have also seen rising interest as prototypical models in quantum information theory, see for example \cite{BravyiKoenig} where disorder effects on quantum memory are studied. 

In Section~\ref{sec:XY} we consider the XY spin chain in random transversal magnetic field and start by discussing results from \cite{HSS}, which establish zero-velocity Lieb-Robinson bounds for this model, see Section~\ref{sec:XYdynloc}. The XY chain is the mathematically simplest model of a quantum spin system because it can be reduced, via the Jordan-Wigner transform, to a system of free Fermions which in turn can be solved by diagonalizing an effective one-particle Hamiltonian. This is reviewed in Section~\ref{sec:XYeffHam}. In the case of the {\it isotropic} XY chain (or XX chain) the one-particle Hamiltonian becomes the one-dimensional Anderson model, where a wealth of localization results are readily available. For the more general {\it anisotropic} XY chain one instead arrives at a less well understood {\it random block operator}. This and other motivations have prompted a considerable amount of recent work on localization for this type of one-particle operators, which will be reviewed in Section~\ref{sec:XYapp}. 

For one-particle random operators it is well known that dynamical localization implies spectral localization. In Section~\ref{sec:xycorrelations} we discuss a result from \cite{HSS} which can be considered as a many-body analogue of this fact: For general classes of quantum spin systems, a zero-velocity Lieb-Robinson bound implies exponential decay of ground state correlations. While we refer to \cite{HSS} for the detailed statement of this general result, Theorem~\ref{thm:XYcordecay} below states what can be obtained in this way for the XY chain.

The other simple quantum many-body model where localization effects due to disorder will be reviewed here are interacting systems of harmonic quantum oscillators, see Section~\ref{sec:LO}, which is based on results from \cite{NSS1} and \cite{NSS2}. Again the associated many-body Hamiltonian can be reduced to an effective one-particle Hamiltonian, including, for a special case, the Anderson model. When compared to the XY chain, the reduction of oscillator systems to one-particle Hamiltonians is even more direct and, in particular, avoids the difficulties in the study of the XY chain arising from the non-locality of the Jordan-Wigner transform.

As a result, for disordered oscillator systems we get a more comprehensive list of localization properties. First, we can deal with oscillator systems on lattices of arbitrary dimension. Also, in addition to dynamical localization in the form of a zero-velocity Lieb-Robinson bound (Section~\ref{sec:dl}) and exponential decay of ground state correlations (Theorem~\ref{thm:gsc} in Section~\ref{sec:oscordecay}), one can also consider correlations of thermal states at arbitrary positive termperature (Theorem~\ref{thm:tsc} in Section~\ref{sec:oscordecay}). Finally, following results in \cite{NSS2}, one can also obtain area laws for ground states as well as for thermal states, see Section~\ref{sec:OSarealaws}. 

What is important about the results on correlation decay and area laws for disordered oscillator systems is that they don't require to assume a uniform ground state gap, and thus, as discussed above, illustrate that the existence of a mobility gap suffices to yield many-body localization properties. 

We conclude this introduction by noting that, as a by-product of trying to better understand many-body localization, the two simple many-body models discussed here have led to various new questions in single-particle random operator theory. In the past, most research in localization theory was focused on the specific type of randomness appearing in the Anderson model. The many-body models discussed here can be randomized in physically well justified ways which lead to types of one particle random operators that are much less well understood. The disordered XY chain has led to the need of better understanding random block operators. Similarly, oscillator systems provide situations where random operators with off-diagonal or multiplicative randomness appear naturally, i.e.\ additional models for which considerably less is known than for the Anderson model.

\bigskip
\noindent
{\em Acknowledgement.}  Very special thanks are due to Jacob Chapman, Eman Hamza and Bruno Nachtergaele for the fruitful collaborations which led the results reviewed here. This work was supported in part by the U.S.\ 
National Science Foundation under grants DMS-1101345 (R.\ S.) and DMS-1069320 (G.\ S.).

\section{The XY Spin Chain in Transversal Field} \label{sec:XY}

The XY quantum spin chain in transversal exterior magnetic field is one of very few exactly solvable models of interacting many-particle quantum systems, as was first understood by Lieb, Schultz and Mattis in their ground breaking work \cite{LSM}. Here, as generally in the theory of quantum spin systems, `particles' are given by spins whose states are described by normalized vectors in {\it finite}-dimensional (in fact, rather low-dimensional) Hilbert spaces. This makes spin systems good models to study effects of interactions, as the underlying non-interacting (or, equivalently, single particle) system is essentially trivial.

The XY chain is a model for $1/2$-spins, meaning that the one-particle space is $\Cx^2$ with canonical basis vectors
\begin{equation} 
e_\uparrow = \left( \begin{array}{c} 1 \\ 0 \end{array} \right) \quad \mbox{and} \quad e_\downarrow = \left( \begin{array}{c} 0 \\ 1 \end{array} \right),
\end{equation}
representing up and down spins. A linear chain of $n$ spins is described by states in the $n$-fold tensor product
\begin{equation} \label{eq:XY}
\H_n = \bigotimes_{j=1}^n \Cx^2
\end{equation}
with orthonormal basis $e_{(j_1,\ldots,j_n)} = e_{j_1} \otimes \ldots \otimes e_{j_n}$, $j_1, \ldots, j_n \in \{\uparrow, \downarrow\}$.

The XY chain is the prototypical model considered as a first example in just about all theoretical investigations of quantum spin systems. This is due to the the fact, observed in \cite{LSM}, that it is described by a Hamiltonian in the $2^n$-dimensional Hilbert space $\H_n$ whose diagonalization can be reduced (via the Jordan-Wigner transform) to that of an effective one-particle Hamiltonian in a $2n$-dimensional Hilbert space.

In its most general form (as long as only next-neighbor interactions are considered) the Hamiltonian of the XY chain in transversal exterior field is given by the self-adjoint operator
\begin{equation} \label{eq:XYchain}
H_n = \sum_{j=1}^{n-1} \mu_j [ (1+\gamma_j) \sigma_j^X \sigma_{j+1}^X + (1-\gamma_j) \sigma_j^Y \sigma_{j+1}^Y] + \sum_{j=1}^n \nu_j \sigma_j^Z
\end{equation}
in $\H_n$. The three sequences of coefficients are assumed to be real, with $\mu_j$ describing the interaction strength, $\gamma_j$ the degree of anisotropy in $X$ and $Y$, and $\nu_j$ the strength of the exterior field. One may also assume $\mu_j \not=0$, as otherwise the chain decomposes into non-interacting shorter pieces. The form (\ref{eq:XYchain}) of the XY chain is the case of open boundary conditions. Periodic boundary conditions, given by an additional interaction between the first and $n$-th spin, could also be considered.

We work with the standard representation of the Pauli martrices
\begin{equation}  \label{pauli}
\sigma^X = \left( \begin{array}{cc} 0 & 1 \\ 1 & 0 \end{array} \right), \quad \sigma^Y = \left( \begin{array}{cc} 0 & -i \\ i & 0 \end{array} \right), \quad \sigma^Z = \left( \begin{array}{cc} 1 & 0 \\ 0 & -1 \end{array} \right), 
\end{equation}
while, essentially, all that matters are their anti-commutation properties
\begin{equation} \label{CAR}
(\sigma^X)^2 = (\sigma^Y)^2 = (\sigma^Z)^2 = I, \quad \{\sigma^X, \sigma^Y\} = \{\sigma^Y, \sigma^Z\} = \{\sigma^Z, \sigma^X\} = 0.
\end{equation}
Finally, the matrix subscripts on the right hand side of (\ref{eq:XYchain}) indicate which component of the tensor product these matrices act.

\subsection{The Effective One-Particle Hamiltonian} \label{sec:XYeffHam}

The effective one-particle Hamiltonian associated with $H_n$ via the Jordan-Wigner transform can be represented as the $2n \times 2n$-block-matrix
\begin{equation} \label{eq:blockoperator}
\tilde{h}_n = \left( \begin{array}{cc} A_n & B_n \\ -B_n & -A_n \end{array} \right),
\end{equation}
acting on ${\mathbb C}^{2n}$, with $n\times n$-matrices
\begin{equation} \label{eq:AandB}
A_n = \left( \begin{array}{cccc} \nu_1 & -\mu_1 & & \\ -\mu_1 & \ddots & \ddots & \\ & \ddots & \ddots & -\mu_{n-1} \\ & & -\mu_{n-1} & \nu_n \end{array} \right), \quad B_n = \left( \begin{array}{cccc} 0 & -\mu_1 \gamma_1 & & \\ \mu_1 \gamma_1 & \ddots & \ddots & \\ & \ddots & \ddots & -\mu_{n-1} \gamma_{n-1} \\ & & \mu_{n-1} \gamma_{n-1} & 0 \end{array} \right).
\end{equation}
For our purposes it is more convenient to reorder the canonical basis vectors $e_1$, $e_2$, \ldots, $e_{2n}$ of ${\mathbb C}^{2n}$ as $e_1$, $e_{n+1}$, \ldots, $e_n$, $e_{2n}$, which turns $\tilde{h}_n$ into the unitarily equivalent $2\times 2$-block Jacobi matrix
\begin{equation} \label{eq:blockJacobi}
h_n = \left( \begin{array}{cccc} \nu_1 J & -\mu_1 S(\gamma_1) & & \\ -\mu_1 S(\gamma_1)^t & \ddots & \ddots & \\ & \ddots & \ddots & -\mu_{n-1} S(\gamma_{n-1}) \\ & & -\mu_{n-1} S(\gamma_{n-1})^t & \nu_n J \end{array} \right),
\end{equation}
where
\begin{equation}
J = \left( \begin{array}{cc} 1 & 0 \\ 0 & -1 \end{array} \right), \quad S(\gamma) = \left( \begin{array}{cc} 1 & \gamma \\ -\gamma & -1 \end{array} \right).
\end{equation}

The matrix $\tilde{h}_n$ (and thus $h_n$) is unitarily equivalent to its negative,
\begin{equation}
\left( \begin{array}{cc} 0 & I \\ I & 0 \end{array} \right) \tilde{h}_n \left( \begin{array}{cc} 0 & I \\ I & 0 \end{array} \right) = -\tilde{h}_n.
\end{equation}
Thus the $2n$ eigenvalues and eigenvectors of $h_n$ are determined by $n$ non-negative eigenvalues $0\le \lambda_1 \le \lambda_2 \le \ldots \le \lambda_n$ and corresponding eigenvectors. 

The Jordan-Wigner transform reduces $H_n$ to a {\it free Fermion system} governed by $h_n$ (or $\tilde{h}_n$). In particular, this means that all $2^n$ eigenvalues and eigenvectors of $H_n$ can be explicitly expressed in terms of the non-negative eigenvalues $\lambda_1$, \ldots, $\lambda_n$ of $h_n$ and the corresponding $n$ eigenvectors. Lieb, Schultz and Mattis \cite{LSM} considered the case of constant coefficients $\mu_m=\mu$, $\gamma_j = \gamma$, $\nu_j = \nu$ for all $j=1,\ldots,n$. In this case the eigenvalues and eigenvectors of $h_n$ can be determined explicitly. Thus, via the Jordan-Wigner transform, $H_n$ becomes exactly solvable (i.e.\ explicitly diagonalizable). Full details of this are reviewed in \cite{HSS}.

For the case of non-constant coefficients the effective Hamiltonian $h_n$ is, in general, not explicitly diagonalizable. However, via ``undoing'' Jordan-Wigner, one may still turn information about $h_n$ into information about $H_n$, and thus reduce the study of a many-body (spin) Hamiltonian to that of an effective one-particle Hamiltonian. This includes information on dynamics, which is our main interest here.

The latter is due to a relation between the unitary dynamics $e^{-itH_n}$ and $e^{-ith_n}$ of the many-body and effective one-particle Hamiltonians. To describe this in more detail, consider the operators
\begin{equation} \label{eq:JordanWigner}
c_1 = a_1, \quad c_j = \sigma_1^Z \ldots \sigma_{j-1}^Z a_j, \quad j=2,\ldots,n,
\end{equation}
introduced in the Jordan-Wigner transform. Here $a_j = (\sigma_j^X -i\sigma_j^Y)/2$.

Then, with $\tau_t^n(A) := e^{itH_n} A e^{-itH_n}$ denoting the many-body Heisenberg dynamics of an operator $A$ on $\H_n$,
\begin{equation} \label{eq:cevolution}
\left( \begin{array}{c} \tau_t^n(c_j) \\ \tau_t^n(c_j^*) \end{array} \right) = \sum_{k=1}^n \; \langle j | e^{-2ith_n} | k \rangle \left( \begin{array}{c} c_k \\ c_k^* \end{array} \right)
\end{equation}
for all $j=1,\ldots,n$. Here we write $\langle j | e^{-2ith_n} | k \rangle$ for the $(j,k)$-th $2\times 2$-block-matrix element of the time evolution of $h_n$, corresponding to the block Jacobi matrix representation of $h_n$ given by (\ref{eq:blockJacobi}). A proof of (\ref{eq:cevolution}) is given in \cite{HSS} (where it is stated in terms of $\tilde{h}_n$).

\subsection{Dynamical Localization} \label{sec:XYdynloc}

Disorder is introduced into the XY chain (\ref{eq:XY}) by considering the coefficients $\mu_j$, $\gamma_j$ and $\nu_j$ as random variables. While more specific assumptions will be made in our later applications, for the following general result we only assume that all coefficients are random variables over a probability space $(\Omega, {\mathcal M}, {\mathbb P})$, with ${\mathbb E}(X) = \int_{\Omega} X\,d\,{\mathbb P}$ denoting expectation. 

\begin{theorem} \label{thm:XYdynloc}
Suppose that $h_n$ is dynamically localized in the sense that there exist $C<\infty$ and $\eta>0$ such that
\begin{equation} \label{eq:1Pdynloc}
{\mathbb E} \left( \sup_{t\in {\mathbb R}} \left\| \langle j | e^{-ith_n} | k \rangle \right\| \right) \le C e^{-\eta |j-k|}
\end{equation}
for all $n\in {\mathbb N}$ and $1\le j,k \le n$.

Then $H_n$ satisfies a zero-velocity Lieb-Robinson bound, i.e.\
\begin{equation} \label{eq:XYdynloc}
{\mathbb E} \left( \sup_{t\in {\mathbb R}} \| [ \tau_t^n(A), B] \| \right) \le C' \|A\| \|B\| e^{-\eta |j-k|}
\end{equation}
for all $1\le j,k \le n$, $A \in {\mathcal A}_j$ and $B\in {\mathcal A}_k$.
\end{theorem}

Here $\eta$ in (\ref{eq:XYdynloc}) is the same constant as in (\ref{eq:1Pdynloc}) and one may choose $C' = 96C/(1-e^{-\eta})^2$. What is important is that the constants do not depend on $n$, thus the many-body dynamical localization bound (\ref{eq:XYdynloc}) is {\it uniform} in the number of spins.

By ${\mathcal A}_j$ we denote the set of local operators acting only on the $j$-th spin, i.e.\ simple tensor product operators acting as the identity away from the $j$-th component. The norms $\|\cdot\|$ appearing in (\ref{eq:1Pdynloc}) and (\ref{eq:XYdynloc}) are any matrix norm in ${\mathbb C}^{2\times 2}$ and the operator norm in $B({\mathcal H}_n)$, respectively.

Going back to \cite{lieb1972}, deterministic Lieb-Robinson bounds of the form
\begin{equation} \label{eq:detLR}
\| [ \tau_t^n(A), B]\| \le C\|A\| \|B\| e^{-\eta (|j-k|-vt)}
\end{equation}
for some $v<\infty$ have been proven for large classes of quantum spin systems, see also \cite{E+Q, hastlec} for recent work on Lieb-Robinson bounds as well as their applications. They can be interpreted as establishing a finite upper bound $v$ on the group velocity or speed of information propagation in the spin system. The bound (\ref{eq:XYdynloc}) is a version of (\ref{eq:detLR}) with $v=0$ (and averaged over the disorder). One may say that in the case of the disordered XY chain the {\it one-particle} dynamical localization bound (\ref{eq:1Pdynloc}) leads to the {\it many-body} dynamical localization bound (\ref{eq:XYdynloc}), expressed in the form of a zero-velocity Lieb-Robinson bound.

For the detailed proof of Theorem~\ref{thm:XYdynloc} we refer to \cite{HSS}. The central ingredient is the relation (\ref{eq:cevolution}). An additional complication comes from the non-local nature of the Jordan-Wigner transform (\ref{eq:JordanWigner}). However, it turns out that in the context considered here it is relatively easy to disentangle this non-locality. Essentially, all that is required is to sum up two geometric series, a trace of which is visible in how the constant $C'$ in (\ref{eq:XYdynloc}) arises from the constants $C$ and $\eta$ in (\ref{eq:1Pdynloc}).

\subsection{Applications} \label{sec:XYapp}

Applications of Theorem~\ref{thm:XYdynloc} lie in identifying parameter regimes in which dynamical localization in the form (\ref{eq:1Pdynloc}) can be shown for the effective one-particle Hamiltonian $h_n$. This is one of the reasons for the recent interest in random block operators of the type (\ref{eq:blockoperator}), as well as of closely related types. Another motivation was provided by attempts to understand disorder effects on solutions of the Bogoliubov-de Gennes equation in BCS theory, see \cite{Kirschetal, GebertMuller, Elgartetal, ElgartSchmidt, Drabkinetal}. 

Random block operators are mathematically interesting because they provide examples of random operators which are not monotone in the random parameters (appearing, e.g., in the form of off-diagonal randomness or of sign changes in the diagonal terms). Therefore the aim of proving dynamical localization results such as (\ref{eq:1Pdynloc}) has led to new challenges in the theory of single-particle random Hamiltonians.

In the following we will entirely focus on applications of Theorem~\ref{thm:XYdynloc} to the case of random exterior field, i.e.\ the parameters
\begin{equation} \label{eq:parameters}
\mu_j = \mu \not= 0 \quad \mbox{and} \quad \gamma_j = \gamma, \quad j=1,2,\ldots
\end{equation}
will be kept constant. The parameters $\nu_j$ characterizing the strength of the exterior transversal field will be chosen as i.i.d.\ random variables. This is the case to which most of the results available in the literature apply. Some remarks on the case of random $\mu_j$ or random $\gamma_j$ can be found in Section~6.4 of \cite{ChapmanStolz}.

\subsubsection{The isotropic XY chain} \label{sec:isoXY}

The most straightforward application of Theorem~\ref{thm:XYdynloc} is the isotropic XY chain in random field, i.e.\ the case $\gamma=0$ in (\ref{eq:parameters}). In this case $B_n=0$ and
\begin{equation} \label{eq:Anderson}
A_n = \left( \begin{array}{cccc} \nu_1 & -\mu & & \\ -\mu & \ddots & \ddots & \\ & \ddots & \ddots & -\mu \\ & & -\mu & \nu_n \end{array} \right)
\end{equation}
is the Anderson model on $\{1,\ldots,n\}$. Thus $\tilde{h}_n$ in (\ref{eq:blockoperator}) is simply the uncoupled direct sum of a positive and a negative Anderson model. In this case the dynamical localization bound (\ref{eq:1Pdynloc}) follows from the corresponding bound
\begin{equation} \label{eq:Anddynloc}
{\mathbb E} \left( \sup_{t\in {\mathbb R}} | \langle \delta_j, e^{-itA_n} \delta_k \rangle | \right) \le C e^{-\eta |j-k|}
\end{equation}
for the Anderson model. The latter is known, for example, if $(\nu_j)_{j=1}^{\infty}$ are i.i.d.\ with absolutely continuous distribution of bounded and compactly supported density. This can be proven either by the Kunz-Souillard method \cite{KunzSouillard} or the fractional moment method (see, e.g., \cite{stolz11} for an introduction). In fact, the fractional moment method can handle substantially weaker assumptions, such as those used in the paper \cite{Elgartetal} which will be discussed in the next subsection.

Thus the isotropic XY chain in random field provided the first example of a disordered quantum system for which the many-body dynamical localization bound (\ref{eq:XYdynloc}) was proven. In somewhat weaker form this had been conjectured in \cite{BurrellOsborne}, partly based on physical heuristics.

\subsubsection{Exterior fields of large disorder} \label{sec:XYlargedisorder}

Important progress in the localization theory of matrix-valued discrete random Schr\"odinger-type operators, including operators of the form $h_n$ and their higher-dimensional analogues, was subsequently made in \cite{Elgartetal}. There dynamical localization bounds of the form (\ref{eq:1Pdynloc}) are established, both in finite and infinite volume, for non-monotone matrix-valued potentials such as $\nu_j J$, $j=1,2,\ldots$, in (\ref{eq:blockJacobi}), assuming large disorder.

More precisely, the result of \cite{Elgartetal} covers our model (\ref{eq:blockJacobi}) in the case where $\mu_j = \mu \not= 0$, $\gamma_j = \gamma$ arbitrary and $\nu_j$ is replaced by $g\nu_j$, $g$ sufficiently large and $\nu_j$, $j=1,2,\ldots$, i.i.d.\ with H\"older-continuous distribution and finite $q$-moment for some $q>0$.

The arguments in \cite{Elgartetal} proceed via an adaptation of the fractional moment method to non-monotone Anderson-type models. In particular, for sufficiently large disorder and some $s>0$ it is shown that the $s$-fractional moments of Green's function decay exponentially, which is then used to deduce dynamical localization.

\subsubsection{The anisotropic XY chain in general random field} \label{sec:generalXY}

The strength of the result in \cite{Elgartetal} lies in its generality and, in particular, in applying to arbitrary dimension. For a one-dimensional model such as (\ref{eq:blockJacobi}), however, it is generally expected that localization should not require large disorder, as other physical mechanisms are at work and other mathematical tools are available.

In \cite{ChapmanStolz} the operator $h_n$ was considered under the following assumptions:
\begin{equation} \label{eq:assump1}
\mu_j=1, \quad \gamma_j = \gamma \in (0,1) \cup (1,\infty) \quad \mbox{for all $j$},
\end{equation}
and
\begin{equation} \label{eq:assump2}
\nu_j, \: j=1,2,\ldots, \:\mbox{are i.i.d.\ random variables with non-trivial distribution of compact support.}
\end{equation}

Of course, other constants $\mu_j = \mu \not=0$ could be considered, requiring only rescaling. The constants $\mu$ and/or $\gamma$ could also be negative (by simple unitary equivalences). The case $\gamma=1$ (and $\mu=1$) corresponds to the Ising model in (\ref{eq:XYchain}) and needs to be treated separately, essentially because the off-diagonal block $S(\gamma)$ in (\ref{eq:blockJacobi}) is not invertible for this case. This will be discussed in Section~\ref{sec:Ising} below. In (\ref{eq:assump2}) compact support of the distribution is assumed mostly for simplicity. This could certainly be generalized, requiring only a suitable decay condition at $\pm\infty$. No regularity of the distribution is required, in particular Bernoulli random variables are allowed. In this case dynamical localization with exponential decay as in (\ref{eq:1Pdynloc}) is not known, not even for the one-dimensional Anderson model. Thus the subexponential decay shown in the following result is the best one can hope to obtain with currently available tools.

\begin{theorem} \label{thm:ChapmanStolz}
Assume that $h_n$ is given by (\ref{eq:blockJacobi}) with coefficients satisfying (\ref{eq:assump1}) and (\ref{eq:assump2}). Then for every compact interval $J\subset {\mathbb R} \setminus \{0\}$ and every $\zeta \in (0,1)$ there are constants $C=C(J,\zeta) < \infty$ and $\eta = \eta(J,\zeta)>0$ such that
\begin{equation} \label{eq:ChapmanStolz}
{\mathbb E} \left( \sup_{t\in {\mathbb R}} \left\| \langle j | e^{-ith_n} \chi_J(h_n) | k \rangle \right\| \right) \le C e^{-\eta |j-k|^{\zeta}}
\end{equation}
for all $n\in {\mathbb N}$ and $j,k \in \{1,\ldots,n\}$.
\end{theorem}

This is shown in \cite{ChapmanStolz}, which is based on the thesis \cite{Chapman}, by methods similar to those which were used in \cite{KleinLacroixSpeis} to prove localization for Anderson models on strips. In particular, the Goldsheyd-Margulis criterion \cite{GoldsheydMargulis} is used to prove positivity of both leading Lyapunov exponents for non-zero energy (note that the block Jacobi matrices (\ref{eq:blockJacobi}) lead to $4\times 4$-transfer matrices and two pairs of sign-symmetric Lyapunov exponents). The paper \cite{ChapmanStolz} also provides a general Thouless formula for ergodic block Jacobi matrices used in the proof. That this leads to subexponential decay in (\ref{eq:ChapmanStolz}) is a consequence of the bootstrap multiscale analysis by Germinet and Klein \cite{GerminetKlein}.

The need for the inclusion of the spectral projection $\chi_J(h_n)$ onto an interval $J\subset {\mathbb R}\setminus \{0\}$ in (\ref{eq:ChapmanStolz}) stems from the fact that at $E=0$ the transfer matrices associated with $h_n$ lose the irreducibility property required by the Goldsheyd-Margulis criterion. Thus, in general, it remains an open question if dynamical localization for $h_n$ can be proven in a vicinity of $E=0$. We will comment on this problem some more in the context of the Ising model below.

However, this problem does not occur if the random exterior field is sufficiently strong. This is due to the following general fact, whose simple proof is provided in \cite{Chapman} (or, in a slightly different setting, in \cite{Kirschetal}): If $\lambda>0$ and $A_n \ge \lambda$ or $-A_n \ge \lambda$, then the block operator (\ref{eq:blockoperator}) has a spectral gap containing $(-\lambda,\lambda)$. 

Moreover, if the support of the distribution of $\nu_j$ is contained in an interval $[a,b]$ such that either $a> 2$ or $b<-2$, then (setting all $\mu_j=1$) the spectrum of $A_n$ given by (\ref{eq:AandB}) is contained in $[2-a,\infty)$ or $(-\infty,-2-b]$,  respectively. Thus $h_n$ has spectral gap $(-\lambda,\lambda)$ with $\lambda = 2-a$ or $\lambda = -2-b$, uniformly in $n$ and in the disorder. Thus, also due to our assumption of boundedness of the $\nu_j$, we can find two compact intervals $J_1$ and $J_2$, both avoiding $E=0$, whose union covers the entire spectrum of $h_n$. Applying (\ref{eq:ChapmanStolz}) to $J_1$ and $J_2$ we get dynamical localization of the entire spectrum for this case,
\begin{equation} \label{eq:dynloczeta}
{\mathbb E} \left( \sup_{t\in {\mathbb R}} \left\| \langle j | e^{-ith_n} | k \rangle \right\| \right) \le C e^{-\eta |j-k|^{\zeta}}.
\end{equation}
While this subexponential decay bound is weaker than the bound (\ref{eq:1Pdynloc}) required in Theorem~\ref{thm:XYdynloc}, one can adjust the proof (see \cite{Chapman}) to show that (\ref{eq:dynloczeta}) implies many-body dynamical localization in the form
\begin{equation} \label{eq:LRsubexp}
{\mathbb E} \left( \sup_{t\in {\mathbb R}} \| [ \tau_t^n(A), B] \| \right) \le C' \|A\| \|B\| e^{-\eta' |j-k|^{\zeta}}.
\end{equation}
In fact, one can choose any $\eta' < \eta$ here.

We mention that there is no obvious general way to turn an energy-localized one-particle dynamical localization bound such as (\ref{eq:ChapmanStolz}) into a many-body dynamical localization bound. The reason is that, essentially, energies of the many-body system are sums of the one-particle energies. Thus, the need of excluding a vicinity of $E=0$ in the one-particle localization result will have global consequences for the spectrum of the many-body system, preventing the deduction of a zero-velocity Lieb-Robinson-type bound.

\subsubsection{The Ising model} \label{sec:Ising}

If $\mu_j = \gamma_j = 1$ for all $j$, then the spin chain (\ref{eq:XYchain}) becomes the quantum Ising model in exterior field,
\begin{equation} \label{eq:Ising}
H_n^{\rm Ising} = 2 \sum_{j=1}^{n-1} \sigma_j^X \sigma_{j+1}^X + \sum_{j=1}^n \nu_j \sigma_j^Z.
\end{equation}
This case was excluded from our earlier discussion because the off-diagonal block
\begin{equation}
S(1) = \left( \begin{array}{cc} 1 & 1 \\ -1 & -1 \end{array} \right)
\end{equation}
of the one-particle operator $h_n^{\rm Ising}$ corresponding to this case via (\ref{eq:blockJacobi}) is not invertible. Thus we can not directly employ a $4\times 4$-transfer matrix formalism. However, a straightforward unitary transformation greatly simplifies the treatment of the Ising case. For this, let us work with the infinite volume version $h_{\infty}^{\rm Ising}$ of the one-particle operator, acting on $\ell^2({\mathbb Z}, {\mathbb C}^2)$. Consider the unitary ${\mathcal U} := \oplus_{j\in {\mathbb Z}} U$ in this space, where
\begin{equation}
U = \frac{1}{\sqrt{2}} \left( \begin{array}{cc} 1 & 1 \\ 1 & -1 \end{array} \right) = U^{-1}.
\end{equation}
Then
\begin{equation} \label{eq:Isingonebody}
h_{\infty}^{\rm Ising} \:\: \cong \:\: {\mathcal U} h_{\infty}^{\rm Ising} {\mathcal U}^{-1} \:\: = \:\: \left( \begin{array}{ccccccc} & \ddots & & & &  & \\ \ddots & \ddots & -2 & & & & \\ & -2 & 0 & \nu_0 & & & \\ & & \nu_0 & 0 & -2 & & \\ & & & -2 & 0 & \nu_1 & \\ & & & & \nu_1 & \ddots & \ddots \\ & & & & & \ddots & \end{array} \right),
\end{equation}
a standard Jacobi matrix ergodic under the 2-shift, which should be easy to analyze. Indeed, for any non-trivial distribution of $\nu_j$ it can be seen by F\"urstenberg's theorem that the leading Lyapunov exponent $\gamma(E)$ is strictly positive for $E\not= 0$. This can be turned into a dynamical localization result similar to Theorem~\ref{thm:ChapmanStolz}, again excluding a vicinity of $E=0$, see \cite{Chapman} for details.

Assume that the support of the distribution of $\nu_j$ is contained in $[a,b]$. There are two cases in which $h_{\infty}^{\rm Ising}$ (and its finite volume versions) have a spectral gap near $E=0$: (i) $[a,b] \subset (2,\infty)$ (equivalently $[a,b] \subset (-\infty,-2)$), or (ii) $[a,b] \subset (0,2)$ (equivalently $[a,b] \subset (-2,0)$). In both cases one gets a one-particle dynamical localization bound of the form (\ref{eq:dynloczeta}), implying a many-body dynamical localization bound of the form (\ref{eq:LRsubexp}) for the Ising model in random field.

However, for more general distribution, $E=0$ may be contained in the almost sure spectrum of $h_{\infty}^{\rm Ising}$. In this case, similar to what was described for the anisotropic XY model in Section~\ref{sec:generalXY}, but in an even more elementary setting, a zero-energy singularity appears which is not yet sufficiently well understood. By the law of large numbers, the Lyapunov exponent at $E=0$ is easily found to be
\begin{equation}
\gamma(0) = \frac{1}{2} \left| {\mathbb E} \left( \log \frac{|\nu_j|}{2} \right) \right|,
\end{equation}
which is non-zero for generic distributions of $\nu_j$. But even in such cases problems remain with completing a proof of dynamical localization. Due to the reducibility of the 2-step transfer matrices of $h_{\infty}^{\rm Ising}$ at $E=0$ (and resulting non-uniqueness of the invariant measure of the Markov chain describing the associated phase dynamics), the usual proof of H\"older continuity of $\gamma(E)$ (e.g.\ \cite{CarmonaLacroix}) fails at $E=0$. 

Finding a proof of H\"older continuity of $\gamma(E)$ at $E=0$ is an interesting open problem, as it would provide the missing step in a proof of many-body dynamical localization, i.e. (\ref{eq:LRsubexp}), for the Ising model (\ref{eq:Ising}) under assumptions on the random field much weaker than the special cases described above. Besides, the surprising fact that open problems remain for a one-particle random operator as simple as (\ref{eq:Isingonebody}) can be considered a challenge by itself. A similar problem is to understand dynamical localization near $E=0$ for the block operators (\ref{eq:blockJacobi}) related to the anisotropic XY chain, in cases other than the simple gapped case described in Section~\ref{sec:generalXY}. This is more difficult than the case of the Ising chain, as it would require a very detailed analysis of the $E$-dependence of the multi-dimensional dynamical systems governing the dynamics of the $4\times 4$-transfer matrices.

\subsection{Decay of Ground State Correlations} \label{sec:xycorrelations}

The zero-velocity Lieb-Robinson bounds discussed so far are a form of many-body {\it dynamical} localization. We will now complement this with a result on {\it eigenvector} localization (as discussed in the introduction, given that Hamiltonians of finite spin systems act in finite-dimensional Hilbert spaces and thus have purely discrete spectrum, in the context considered here it makes no sense to consider spectral localization). Thus we ask if eigenvectors of disordered interacting many-body systems are, in suitable sense, close to product states, the eigenvectors of non-interacting many-body systems.

A first (although potentially unreliable) indication of closeness to product states can be found by considering correlations of local operators (e.g.\ $A \in {\mathcal A}_j$ and $B\in {\mathcal A}_k$ as  in Section~\ref{sec:XYdynloc}, $|j-k|$ large) in such states. For product states all such correlations vanish.

We focus on the isotropic XY chain in random exterior field, with assumptions as in Section~\ref{sec:isoXY}, i.e.\ the $\nu_j$ are i.i.d.\ with bounded and compactly supported density. The following result on exponential decay of ground state correlations was shown in \cite{HSS}. 

\begin{theorem} \label{thm:XYcordecay}
Under the assumptions of Section~\ref{sec:isoXY} the Hamiltonian $H_n$ in (\ref{eq:XYchain}) almost surely has a unique normalized ground state $\psi_0$ and there exist constants $C<\infty$ and $\eta>0$ such that
\begin{equation} \label{eq:XYcordecay}
{\mathbb E} \left( \left| \langle \psi_0, AB \psi_0 \rangle - \langle \psi_0, A \psi_0 \rangle \langle \psi_0, B \psi_0 \rangle \right| \right) \le \|A\| \|B\| \,n\, e^{-\eta |j-k|}
\end{equation}
for all $A \in {\mathcal A}_j$ and $B\in {\mathcal A}_k$ and all $1\le j < k \le n$.
\end{theorem}

Much earlier it had been argued by Klein and Perez in \cite{KleinPerez} that correlations between the local spin raising and lowering operators $a_j^* = (\sigma_j^X +i \sigma_j^Y)/2$ and $a_k = (\sigma_k^X - i\sigma_k^Y)/2$ decay exponentially in the sense that
\begin{equation} \label{eq:KleinPerez}
\sup_n |\langle \psi_0, a_j^* a_k \psi_0 \rangle | \le C_{\nu} e^{-\eta|j-k|}
\end{equation}
for almost every choice $\nu = (\nu_j)$ of the magnetic field. Using explicit calculations for the XY chain, including Wick's theorem, they expand the ground state correlations in terms of the Fermi two-point function $\langle \psi_0, c_j^* c_k \psi_0 \rangle$ (with $c_j$ from (\ref{eq:JordanWigner})).

In \cite{HSS} a different approach is taken by first proving a deterministic result which holds for very general quantum spin systems: A zero-velocity Lieb-Robinson bound such as (\ref{eq:XYdynloc}) implies exponential decay of ground state correlations under mild assumptions on the ground state gap. For the exact statement of this result we refer to Section 2 of \cite{HSS}. The proof follows a strategy developed earlier in \cite{nachtergaele2005d, hastings2005}, where is was shown that deterministic Lieb-Robinson bounds of the form (\ref{eq:detLR}) (i.e.\ with {\it positive} velocity) imply exponential decay of ground state correlations if the ground state gap $E_1-E_0$ of the spin system has a positive lower bound, uniform in the system size. In contrast, the deterministic result proven in \cite{HSS} says that under the stronger assumption of a {\it zero}-velocity Lieb-Robinson bound one can weaken the assumption on the spectral gap. The gap may shrink to zero for increasing system size, leading to an at most logarithmic correction to the exponential decay of ground state correlations.

Theorem~\ref{thm:XYcordecay} above and is proof (in Section~4.2 of \cite{HSS}) are an adaptation of this deterministic result to the setting of the disordered isotropic XY chain. Specifically, it is used that in this case the ground state gap is given by
\begin{equation} \label{eq:gsgap}
E_1 - E_0 = 2\, {\rm dist}(0,\sigma(A_n)),
\end{equation}
where, as before, $A_n$ is the Anderson model on $\{1,\ldots,n\}$. Smallness of the right hand side of (\ref{eq:gsgap}) is controlled by the well known Wegner estimate for the Anderson model, which gives that $E_1-E_0$ can be very small, but that this has small probability. This is the source for the correction factor $n$ on the right hand side of (\ref{eq:XYcordecay}). It is an interesting open problem to understand if this factor can be eliminated, as this would be necessary to draw conclusions for the limiting case of an infinite chain.

We conclude this section with several additional remarks.

While we haven't checked the details, it is likely that the recent work on Wegner estimates for random block operators in \cite{ElgartSchmidt} will allow to extend Theorem~\ref{thm:XYcordecay} to the anisotropic XY chain, at least for the case of large disorder considered in \cite{Elgartetal} and described in Section~\ref{sec:XYlargedisorder} above.

The disordered XY chain should allow to illustrate other, stronger forms of many-body eigenfunction localization in addition to the exponential decay of ground state correlations, e.g. as established by Theorem~\ref{thm:XYcordecay} or the bounds proven in \cite{KleinPerez}. First, one should try to prove exponential decay of correlations for states other than the ground state, at least for low energy states. Of particular interest would be to look at the mixed states given by thermal states, which, due to the presence of disorder, might exhibit exponential correlation decay at any positive temperature. 

A better way than correlation decay to measure closeness of states to product states is to consider {\it entanglement entropy}. Here one might ask if the bipartite entanglement entropy of the ground state of the disordered XY chain satisfies an {\it area law}, which has been argued to hold for general {\it gapped} one-dimensional spin systems in \cite{hast07}. More recently it has been established in \cite{BrandaoHorodecki2012,Brand12} that for one-dimensional systems exponential decay of correlations of a state imply an area law, without assuming a gap. While these works are for deterministic systems, one can hope that their arguments allow to draw similar conclusions for random systems such as the disordered XY chain. 

For the disordered XY chain a more direct analysis of entanglement entropies should be possible, exploiting its reduction to a free Fermion system via the Jordan-Wigner transform (as done in \cite{KleinPerez} to study ground state correlations). However, the non-locality of the Jordan-Wigner transform causes considerable difficulties in attempting this. In the deterministic setting substantial progress on this problem has recently been made in \cite{Bernigauetal}, where general free Fermionic lattice systems are considered and applications to the XY chain are discussed as a special case, assuming a spectral gap. It would be worthwhile to find out if this work can be applied to the random non-gapped case.

\section{Lattice Oscillators} \label{sec:LO}

Another well-studied and immanently analyzable class of quantum models are the lattice oscillators. 
In contrast to the XY chains discussed in the previous section, these multi-dimensional oscillators 
correspond to 'particles' with a continuous degree of freedom. A similarity between these models
is that information on the many-body oscillator dynamics, as well as other quantities of interest, can also be reduced to that of 
an effective single particle Hamiltonian. This key fact is crucial in all of the results reviewed below.

We will consider a class of lattice oscillators defined on $\mathbb{Z}^d$. 
Fix an integer $L \geq 1$ and denote by $\Lambda_L = [-L,L]^d \cap \mathbb{Z}^d$ the cubic set of lattice points, 
centered at the origin, with side length $2L+1$. To each such finite volume $\Lambda_L$, associate the
Hilbert space
\begin{equation} \label{hilbert}
\mathcal{H}_L = \bigotimes_{x \in \Lambda_L} L^2( \mathbb{R}, d q_x) = L^2 \left( \mathbb{R}^{\Lambda_L}, dq \right) 
\end{equation}
where by $q = (q_x)_{x \in \Lambda_L}$ we denote the spatial variables. Here, the choice of one spatial dimension 
in the single-site Hilbert space, i.e. the fact that we use $L^2( \mathbb{R})$ at each $x$ rather than $L^2( \mathbb{R}^{\nu})$, is mainly for 
notational convenience; lattice oscillators over $L^2( \mathbb{R}^{\nu})$ can be analyzed in a manner nearly identical to what is discussed below. 

One commonly studied class of oscillator models, also called harmonic, can be described in terms of three sequences of real
parameters: the masses $\{ m_x \}$ - typically positive, the spring coefficients $\{ k_x \}$, and the couplings $\{ \lambda_{x,y} \}$ -
the latter two generally non-negative. Given these parameters, a finite volume harmonic Hamiltonian $H_L$, acting on
$\mathcal{H}_L$, is given by 
\begin{equation} \label{oscham}
H_L = \sum_{x \in \Lambda_L} \left( \frac{1}{2 m_x} p_x^2 + \frac{k_x}{2} q_x^2 \right) + \sum_{\langle x, y \rangle} \lambda_{x,y} (q_x-q_y)^2 
\end{equation}
where the final sum above is over nearest-neighbor sites $x,y \in \Lambda_L$, i.e. those $x$ and $y$ with $|x-y|=1$ in the 1-norm.
Here $H_L$ describes a d-dimensional system of one-dimensional oscillators that are coupled by nearest 
neighbor quadratic interactions. In (\ref{oscham}), we have denoted by $q_x$ and $p_x$ the position and momentum operators respectively, i.e.
by $q_x$ we mean the operator of multiplication by $q_x$, a slight abuse of notation, and by $p_x$ we mean the
operator $- i \frac{\partial}{\partial q_x}$. It is well-known that each of these operators is self-adjoint on $\mathcal{H}_L$
and, moreover, they satisfy the commutation relations
\begin{equation} \label{CCR}
[p_x, p_y] = [q_x, q_y] = 0 \quad \mbox{and} \quad [q_x,p_y] = i \delta_{x,y} \idty 
\end{equation}
which is the analogue of (\ref{CAR}) in this setting. Thus, in particular, $H_L$ is a well-defined 
self-adjoint, in fact non-negative, operator on $\mathcal{H}_L$ (more precisely, the unbounded operators $q_x$, $p_x$ and $H_L$ can be introduced as the closures of the corresponding operators defined on $C_0^{\infty}(\mathbb{R}^{\Lambda_L})$ or the Schwarz space on $\mathbb{R}^{\Lambda_L}$). This is not, of course, the most general
harmonic model one could consider, but it is sufficiently general to allow us to discuss all of the 
results we wish to review; see the end of Section~\ref{sec:dl} for a further comment in this direction.

At this point, it is interesting to pause and compare the two models being considered. 
In regards to the underlying Hilbert spaces, it is clear that the distinction between (\ref{eq:XY}) and (\ref{hilbert}) 
lies crucially in the fact that we have replaced a finite dimensional Hilbert space, i.e. $\mathbb{C}^2$,
with the infinite dimensional space $L^2( \mathbb{R})$. Moreover, the objects of interest for these oscillator systems are
the unbounded self-adjoint operators, $p_x$ and $q_x$, in contrast to the bounded spin matrices, e.g. as in (\ref{pauli}),
for the XY chain. Although these differences are stark, there is enough similarity to allow for a proof of
analogous results through different techniques, as will be
discussed below. 

Central to the analysis of these oscillator models is their quadratic structure.
By writing  $q =(q_x)_{x \in \Lambda_L}$ and $p =(p_x)_{x \in \Lambda_L}$ as vectors with operator-valued  
components, it is easy to see that 
\begin{equation} \label{baseham}
H_L = (q^T, p^T) \left( \begin{array}{cc} h_L & 0 \\ 0 & \mu \end{array} \right) \left( \begin{array}{c} q \\ p \end{array} \right)
\end{equation}
the latter to be understood as a formal matrix product with $q^T$ and $p^T$ the corresponding row vectors. 
Here $\mu$ is the diagonal matrix with entries $\mu_{x,y} = \frac{1}{2 m_x} \delta_{x,y}$ and the coefficient matrix 
$h_L$ is easily seen to satisfy
\begin{equation}
\langle f, h_L  g \rangle = \sum_{<x,y>} \lambda_{x,y} \overline{(f(x) - f(y))}(g(x)-g(y)) + \sum_{x \in \Lambda_L} \frac{k_x}{2} \overline{f(x)}g(x) 
\end{equation}
for any $f, g \in \ell^2( \Lambda_L)$. It is also well-known, and discussed in detail e.g.\ in \cite{NSS1}, that the spectrum of $H_L$, as well as the corresponding eigenfunctions, can be completely characterized in terms of $h_L$. For this reason, we refer to $h_L$ as an effective single-particle Hamiltonian. An important final observation is that: if the spring coefficients are chosen at random while the masses and couplings
are kept constant, then $h_L$ is precisely the well-studied Anderson model on $\ell^2(\Lambda_L)$; compare e.g. with (\ref{eq:Anderson}).

\subsection{Dynamical Localization} \label{sec:dl}

For deterministic oscillator models, locality estimates, in the form of 
Lieb-Robinson bounds, were only recently proven in \cite{NRSS}. These estimates
demonstrate an upper bound, usually dubbed the system's Lieb-Robinson velocity, 
on the rate at which information can propagate in oscillator models. 
In the context of random oscillators, a zero-velocity Lieb-Robinson bound has been
established in \cite{NSS1} for a certain class of models, indicating a strong form of dynamical
many-body localization. We briefly review these results below.

To understand these locality estimates more precisely, we first introduce the many-body oscillator 
dynamics and a well-chosen class of bounded operators. We will denote by $\mathcal{B}(\mathcal{H}_L)$ 
the set of all bounded linear operators over $\mathcal{H}_L$.
Self-adjointness of the Hamiltonian $H_L$, as in (\ref{oscham}), guarantees the existence of the corresponding Heisenberg dynamics, or time evolution, i.e.
\begin{equation}
\tau_t^L(A) = e^{i t H_L} A e^{-it H_L} \quad \mbox{for any } A \in \mathcal{B}(\mathcal{H}_L) \quad \mbox{and} \quad t \in \mathbb{R}
\end{equation}
For the harmonic models we are considering, a natural class of bounded operators whose dynamical evolution can be readily
investigated are the strictly local Weyl operators.
These are, for any $x \in \Lambda_L$ and $z \in \mathbb{C}$ given by the unitary operators 
\begin{equation}
W_x(z) = {\rm exp} \left[ i \left( {\rm Re}[z] q_x + {\rm Im}[z] p_x \right) \right] \in \mathcal{B}(\mathcal{H}_L) 
\end{equation}
As the position and momentum operators associated to distinct lattice points commute, it is clear that
\begin{equation}
\left[ W_x(z), W_y(z') \right] = 0 \quad \mbox{whenever} \quad x \neq y
\end{equation}
When $t \neq 0$, it is no longer the case that $\tau_t^{L}(W_x(z))$ generally commutes with $W_y(z')$ when $y \neq x$. 
For oscillator models, the analogue of Lieb-Robinson bounds, which are well-known in the context of quantum spin systems, 
quantify this phenomenon.

In \cite{NRSS}, the deterministic, constant-coefficient oscillator model, i.e.  (\ref{oscham}) with $m_x = m >0$, $k_x = k \geq 0$, and $\lambda_{x,y} = \lambda >0$ for all $x \in \mathbb{Z}^d$ and all nearest neighbor pairs $\{x, y\} \subset \mathbb{Z}^d$, was considered.
It was proven that for any $\eta >0$, there are $C< \infty$ and $v>0$ such that
\begin{equation} \label{osclrb}
\| [ \tau_t^L(W_x(z)), W_y(z') ] \| \leq C |z| |z'| \cdot e^{- \eta(|x-y| - v|t|)}
\end{equation}
for all $L$, $x$, $y$, $z$, $z'$, and $t$. 
Hence, the above bound demonstrates that, at least for this model, there is a positive, volume-independent velocity $v$, which is called the
Lieb-Robinson velocity, bounding the rate at which information propagates through this oscillator system. To avoid unnecessary technicalities, we have only stated the above bound in this very specific case, see \cite{NRSS} and \cite{CSE} for analogous estimates applying to more general harmonic models as well as a 
larger class of observables. Moreover, it is shown in \cite{NRSS}, see also \cite{Amour, NSSSZ}, that this
type of bound is robust even under some anharmonic perturbations. 

If the coefficients determining $H_L$ are random and {\it large} - to be understood in an appropriate sense, it is natural to 
expect that a stronger form of locality, stronger than the bounds in (\ref{osclrb}), can be observed. This is what has been proven in
\cite{NSS1}; in fact, a zero-velocity Lieb-Robinson bound is established for certain random oscillator models. 

Let us first restrict our attention to a specific random model. Consider the oscillator model, as in (\ref{oscham}), defined with
constant masses and couplings, i.e., take 
\begin{equation} \label{constm+c}
m_x = 1/2 \quad \mbox{for all } x \in \mathbb{Z}^d \quad \mbox{and} \quad \lambda_{x,y} = 1 \quad \mbox{for all nearest neighbor pairs } \{x, y\}  \subset \mathbb{Z}^d
\end{equation}
Introduce randomness by taking the spring coefficients random with the form
\begin{equation} \label{randsc}
k_x = \mu \, \omega_x \quad \mbox{for all } x \in \mathbb{Z}^d 
\end{equation}
where $\mu >0$ is a disorder parameter and $\{ \omega_x \}$ is an i.i.d.\  sequence of non-negative random variables on a 
probability space $( \Omega, \mathcal{F}, \mathbb{P})$ with $\mathbb{E}(X) = \int_{\Omega}X d \mathbb{P}$ denoting the corresponding
expectation. Here $\mu >0$ large will correspond to large disorder. As mentioned just before the beginning of this section, for this choice of coefficients, the
effective single-particle, finite-volume Hamiltonian $h_L$ is precisely the Anderson model on $\ell^2( \Lambda_L)$, and there is
a wealth of localization literature available for this model. In fact, to be concrete, we will make the
following assumption on these random variables:

{\bf Assumption A:} {\it We will say that the oscillator model $H_L$ in (\ref{oscham}), and also 
the corresponding single-particle Hamiltonian $h_L$, satisfy Assumption A if 
\newline i) both (\ref{constm+c}) and (\ref{randsc}) are satisfied, and 
\newline ii) the common distribution of the non-negative random variables in (\ref{randsc}) 
is absolutely continuous with respect to Lebesgue measure with a
density that is bounded and of compact support in $[0, \infty)$.}

 The following result is proven in \cite{NSS1}.

\begin{theorem} \label{thm:sdl} Suppose that the oscillator model $H_L$, as in (\ref{oscham}), is
random and satisfies Assumption A. Then, for $\mu>0$ sufficiently large, there exists
$\eta >0$ and $C < \infty$ for which
\begin{equation} \label{sdl}
\mathbb{E} \left( \sup_{t \in \mathbb{R}} \left\| \left[ \tau_t^L(W_x(z)), W_y(z') \right] \right\| \right) \leq C |z| |z'| e^{- \eta |x-y|} 
\end{equation}
for all $L$, $x$, $y$, $z$, and $z'$.
\end{theorem}

A few comments are in order. First, in comparison with the proto-typical oscillator locality bound, i.e. (\ref{osclrb}) above, it is clear that the
result in (\ref{sdl}) demonstrates that, upon averaging, the velocity of information propagation in these random models is identically zero. For this reason,
we refer to this result as establishing a zero-velocity Lieb-Robinson bound for this model, which we again stress is a strong form of dynamical, many-body localization.    
Next, the proof of this result requires that the entire spectrum of the single-particle Hamiltonian is localized, i.e. large $\mu>0$ is necessary. For those familiar with the single-particle theory of Anderson localization in one dimension, it is natural to expect that, in the special case $d=1$, Theorem~\ref{thm:sdl}
holds for arbitrary disorder, i.e. for any $\mu >0$. This result is contained in \cite{NSS1}. 

We now outline the two key elements used in proving Theorem~\ref{thm:sdl}.
The first is an important deterministic fact which is not only useful in analyzing harmonic models, but also 
explains the specific class of bounded operators we are considering.
In words, this well-known observation is that the harmonic evolution of a (strictly local) Weyl operator 
is an explicit (non-local) Weyl operator. For example, given any $x \in \Lambda_L$ and $z \in \mathbb{C}$,
there is a function $f_t^x : \Lambda_L \to \mathbb{C}$ for which $f_0^x(y) = z \delta_x(y)$ and
\begin{equation} \label{weylevo}
\tau_t^L(W_x(z)) = {\rm exp} \left[ i \sum_{y \in \Lambda_L} \left( {\rm Re}[f_t^x(y)] q_y + {\rm Im}[f_t^x(y)] p_y \right) \right] \quad \mbox{for all } t \in \mathbb{R}
\end{equation} 
see e.g.\  \cite{NRSS, NSS1} for more details and, in particular, an explicit formula for $f_t^x$. Combining this fact with 
some known properties of Weyl operators, the following bound, a special case of formulas (3.19) and (3.21) of \cite{NSS1}, results from a direct calculation  
\begin{eqnarray} \label{dynbd}
\left\| \left[ \tau_t^L(W_x(z)), W_y(z') \right] \right\| & \leq &  |z| |z'| \left\{ 2 | \langle \delta_x, \cos(2t h_L^{1/2} ) \delta_y \rangle|  + \right. \\
& \mbox{ } & \left. \quad + | \langle \delta_x, h_L^{1/2} \sin(2t h_L^{1/2} ) \delta_y \rangle| +  | \langle \delta_x, h_L^{-1/2} \sin(2t h_L^{1/2} ) \delta_y \rangle|  \right\} \nonumber 
\end{eqnarray}
An illustrative similarity to Theorem~\ref{thm:XYdynloc} on the XY chain is now clear: bounds on the evolution of the effective single-particle Hamiltonian $h_L$
can be used to control the many-body dynamics associated to $H_L$, at least as far as these Lieb-Robinson commutators are
concerned.

The other crucial element in the proof of Theorem~\ref{thm:sdl} focuses on the form of
the bound appearing in (\ref{dynbd}) and some well-known techniques in the single-particle
theory of Anderson localization. We state this as a lemma directly for the Anderson model $h_L$, the proof of
which is contained in \cite{NSS1}.

\begin{lemma} \label{lem:sec}
Let $h_L$ be the $d$-dimensional Anderson model satisfying Assumption A.
If $\mu >0$ is sufficiently large, 
then for every $\alpha > -1$, there exists $C < \infty$ and $\eta >0$ such that
\begin{equation} \label{sec}
\mathbb{E} \left( \sup_{|g| \leq 1} | \langle \delta_x, h_L^{\alpha} g(h_L) \delta_y \rangle | \right) \leq C e^{- \eta |x-y|}
\end{equation}
for all $L$ and $x,y \in \Lambda_L$. 
\end{lemma}
 
The lemma above shows that, on average, a large class of functions of $h_L$ have
matrix elements  that decay exponentially with a rate that is uniform in the length scale $L$. The case of $\alpha =0$ is a commonly studied object
in the theory of single-particle Anderson localization. For example, it is well-known that these {\it eigenfunction correlators}
are a useful tool in deriving results on dynamical localization for the Anderson model, see e.g. \cite{aiz94, aizetal01, stolz11}.
Here, in order to bound the quantities that arise in the study of the many-body oscillator dynamics, e.g.\ in (\ref{dynbd}), 
it becomes necessary to study {\it singular eigenfunction correlators}, and this observation, as well as the bound, 
appears to be new. Bounds of the form (\ref{sec}) for the multi-dimensional Anderson model can currently only be proven by the fractional moments method, which is the reason for requiring strong regularity of the distribution of the $k_x$, as in Assumption A ii).

The proof of Theorem~\ref{thm:sdl} now readily follows. One uses (\ref{dynbd}) and 
applies Lemma~\ref{lem:sec} to the three resulting terms. Each term is bounded similarly, e.g.\ in 
the most singular one corresponding to $\alpha = -1/2$, the estimate is a consequence of the simple fact 
that the functions $g_t(x) = \sin( 2t \sqrt{x})$ clearly satisfy $|g_t| \leq 1$. 

Let us now briefly discuss some generalizations. More general oscillator models which, in the finite volume, have the form 
\begin{equation} \label{genham}
\tilde{H}_L = (q^T, p^T) \left( \begin{array}{cc} h_1 & 0 \\ 0 & h_2 \end{array} \right) \left( \begin{array}{c} q \\ p \end{array} \right)
\end{equation}
can easily be considered; compare with (\ref{baseham}). For random models, one would regard the entries of the coefficient matrices
$h_1$ and $h_2$ as random variables on some probability space $(\Omega, \mathcal{F}, \mathbb{P})$, similarly to 
what was done before. If the matrices $h_1$ and $h_2$ are almost surely positive definite,  then  
\begin{equation}
\tilde{h}_L = h_2^{1/2} h_1 h_2^{1/2}
\end{equation} 
is known to correspond to an effective single-particle Hamiltonian for $\tilde{H}_L$ in the sense described above.
Deterministically, on this almost sure set, the analogue of (\ref{weylevo}), and thereby a variant of (\ref{dynbd}), holds.
It is then clear that an analogue of Theorem~\ref{thm:XYdynloc} holds for these random oscillator models. In other
words, one can pose conditions on the decay of the singular eigenfunction correlators corresponding to $\tilde{h}_L$
under which $\tilde{H}_L$ satisfies a zero-velocity Lieb-Robinson bound. Verifying the analogue of Lemma~\ref{lem:sec}
for more general $\tilde{h}_L$ is an interesting open problem. In fact, in \cite{NSS1} we focused almost exclusively on the
simplest case of random spring coefficients so that the corresponding single-particle Hamiltonian was the Anderson model.
There are already interesting open problems of this type corresponding to cases where just the masses and/or the couplings are randomized. 

Lastly, we have also verified a zero-velocity Lieb-Robinson bound for the position and momentum operators.
For example, the analogue of Theorem~\ref{thm:sdl} holds with $W_x(z)$ replaced by $q_x$ and
$W_y(z')$ replaced by $p_y$. We have chosen to focus on results for Weyl operators in this review to avoid unnecessary
complications due to the fact that $q_x$ and $p_y$ are unbounded operators, see \cite{NSS1} for more details. 

\subsection{Correlation Decay} \label{sec:oscordecay}

An alternative way in which one can attempt to establish many-body localization for a given random
quantum model is to analyze its eigenstates. For general non-interacting quantum systems, 
the eigenstates are simply product states. In this case, it is clear that correlation functions 
in these states factorize. One possibility for indicating that a random, many-body system with
non-trivial interactions is in a localized phase is to show that the correlations, in the
ground state for example, almost factorize - in the sense that they decay exponentially 
when applied to observables with large spatial separation. In \cite{NSS1}, we proved
exponential decay of correlations in both ground states and thermal states for certain random oscillator 
models, as we will review below. 

Although a number of the deterministic calculations discussed below also apply in the case of
general oscillator models, see e.g. those defined by (\ref{genham}), in this section we will mainly restrict 
our attention to the case of the random model defined by (\ref{constm+c}) and (\ref{randsc}) and satisfying Assumption A.
In fact, it is to this specific model that most of the localization results on correlation decay found in \cite{NSS1}
actually apply. 

Before we state the results that we have for random systems, it is
useful to review what is known for deterministic oscillator models.
Consider the oscillator Hamiltonian $H_L$, as in (\ref{oscham}), with uniformly bounded masses, couplings, and
spring coefficients. If the masses additionally satisfy a positive (uniform) lower bound and the single site Hamiltonian
is positive definite, then it is easy to check that the oscillator Hamiltonian $H_L$ has a unique ground state, see e.g.\ \cite{NSS1}.   
Let us denote by $\Omega_L \in \mathcal{H}_L$ the non-degenerate, 
normalized ground state of $H_L$.  In addition, for any $A \in \mathcal{B}(\mathcal{H}_L)$, 
label the ground state expectation of $A$ by
\begin{equation} \label{gse}
\langle A \rangle_L = \langle \Omega_L, A \Omega_L \rangle
\end{equation}

It is well-known that if the oscillator model $H_L$ is gapped, then the ground state
correlations decay exponentially. One typical result of this type is as follows.

\begin{theorem} \label{thm:gscd} Suppose that $m_x = m >0$, $\lambda_{x,y} = \lambda >0$, and that 
$0<a \leq k_x \leq b< \infty$ for all $x \in \mathbb{Z}^d$ and all nearest neighbor pairs $\{x,y \} \subset \mathbb{Z}^d$. 
Then there exist $C < \infty$ and $\eta >0$ such that
\begin{equation} \label{gsd}
\left| \langle W_x(z) W_y(z') \rangle_L - \langle W_x(z) \rangle_L \langle W_y(z') \rangle_L \right| \leq C|z| |z'| e^{- \eta |x-y|} 
\end{equation}
for all $L$, $x$, $y$, $z$, and $z'$. 
\end{theorem}

The key point here is that the lower bound on the spring coefficients $k_x$ ensures that this specific harmonic model is gapped,
i.e.\ the distance between the energies associated to the ground state and the first excited state of  $H_L$ satisfies
$E_1(L) - E_0(L) \geq \gamma >0$ uniformly in $L$. See \cite{SCW, CE, CSE} for generalizations of this result to a 
wide range of deterministic gapped harmonic systems. Similar results also hold for some gapped, anharmonic models, see e.g. \cite{NRSS}. 
Generally, for deterministic oscillator models the decay rate, e.g.\ $\eta = \eta(\gamma)$ in (\ref{gsd}) above, vanishes as the gap goes to zero. 

In the case of random oscillator models, the following result is proven in \cite{NSS1}.

\begin{theorem} \label{thm:gsc} Suppose that the oscillator model $H_L$, as in (\ref{oscham}), is
random and satisfies Assumption A.
Then there exist $C < \infty$ and $\eta >0$ such that
\begin{equation} \label{sgscd}
\mathbb{E} \left( \left| \langle W_x(z) W_y(z') \rangle_L - \langle W_x(z) \rangle_L \langle W_y(z') \rangle_L \right| \right)  \leq C|z| |z'| e^{- \eta |x-y|} 
\end{equation}
for all $L$, $x$, $y$, $z$, and $z'$. In addition, if $\mu>0$ is sufficiently large, then one also has that there exist
$C'< \infty$ and $\eta'>0$ such that 
\begin{equation} \label{dgscd}
\mathbb{E} \left( \sup_{t \in \mathbb{R}} \left| \langle \tau_t^L \left( W_x(z) \right) W_y(z') \rangle_L - \langle W_x(z) \rangle_L \langle W_y(z') \rangle_L \right| \right)  \leq C' |z| |z'| e^{- \eta' |x-y|} 
\end{equation}
for all $L$, $x$, $y$, $z$, and $z'$.
\end{theorem}

Some comments are in order. First, the static result, i.e.\ (\ref{sgscd}) above, which we stress applies in any dimension, 
does not require large disorder, i.e.\ large $\mu >0$. This is due to the fact that, for arbitrary non-trivial 
disorder, the Anderson model is localized at the bottom of its spectrum. 
Moreover, the bound in (\ref{dgscd}) demonstrates that, in the large disorder regime, 
even the dynamically evolved ground state correlations remain localized uniformly in $t$.

Next, it is important to note that the above theorem applies even in some cases where the
oscillator Hamiltonian $H_L$ is not uniformly gapped. For example, suppose that the 
support of the density of the random variables under consideration is $[0, k_{\rm max}]$,
in particular, the support contains a neighborhood of $0$. Under this additional assumption, one
can easily check that almost surely the corresponding oscillator model is gapless in the thermodynamic limit, as there are sufficiently large regions in ${\mathbb Z}^d$ where the values of all $k_x$ are arbitrarily close to $0$.
Yet the correlation decay, as in (\ref{sgscd}) and (\ref{dgscd}), still holds. 
Here, even though the model is gapless, the spectrum just above the ground state energy is {\it localized}.
In the physics literature, it is said that the states corresponding to these energies do not
facilitate mobility, and thus such a model is said to exhibit a {\it mobility gap}.

The proof of Theorem~\ref{thm:gsc} uses well-known, deterministic formulas for the ground state expectation of Weyl operators.
These formulas are given in \cite{NSS1} and, at least in the static case, they readily yield the
bound
\begin{equation} \label{gscbd}
 \left| \langle W_x(z) W_y(z') \rangle_L - \langle W_x(z) \rangle_L \langle W_y(z') \rangle_L \right| \leq  \frac{1}{2} |z| |z'|  \left( | \langle \delta_x, h_L^{-1/2} \delta_y \rangle | + | \langle \delta_x, h_L^{1/2} \delta_y \rangle |\right) 
\end{equation}
Note that in cases where the model is gapless, the first term which appears on the right-hand-side of 
(\ref{gscbd}) is genuinely singular, in contrast to the final term in (\ref{dynbd}) which is, in fact, bounded.
To complete the proof of (\ref{sgscd}) in Theorem~\ref{thm:gsc}, we apply the following lemma from \cite{NSS1}
which does not require an assumption on large disorder.

\begin{lemma} \label{lem:anymu} Let $h_L$ be the d-dimensional Anderson model satisfying Assumption A.
Suppose that $\phi: (0, \infty) \to \mathbb{C}$ satisfies $| \phi(t) | \leq C t^{\alpha}$ for $t$ near $0$ with some
$\alpha >-1$, and that $\phi$ has an analytic extension to a semi-strip $\{ z \in \mathbb{C} : {\rm Re}[z] >0,  |{\rm Im}[z]| \leq \xi \}$
for some $\xi>0$. Then there exist $C'< \infty$ and $\eta >0$ such that
\begin{equation}
\mathbb{E} \left( \left| \langle \delta_x, \phi(h_L) \delta_y \rangle \right| \right) \leq C' e^{- \eta |x-y|} 
\end{equation}
for all $L$ and $x,y \in \Lambda_L$.
\end{lemma}

The proof of Lemma~\ref{lem:anymu} uses that, for any $\mu>0$, the spectrum of $h_L$ near $0$ is localized,
and hence a contour integration argument, similar to methods introduced in \cite{AG}, applies; 
for details see the proof of Proposition A.3(c) in \cite{NSS1}. 

The bound (\ref{sgscd}) now follows by combining (\ref{gscbd}) with Lemma~\ref{lem:anymu}, noting that $\phi_1(t)=t^{-1/2}$ and $\phi_2(t) = t^{1/2}$ satisfy the assumptions required there. We have stated Lemma~\ref{lem:anymu} in this somewhat 
general form so that it is more easily applied in other situations, see e.g.\ comments immediately after Theorem~\ref{thm:tsc} below. 
The remaining dynamical bound, i.e.\ (\ref{dgscd}) in Theorem~\ref{thm:gsc}, follows similarly; a direct
 calculation establishes an estimate analogous to (\ref{gscbd}) and an application of Lemma~\ref{lem:sec} finishes the proof. 

There are, of course, other states in which one can consider correlation decay. 
Statements analogous to Theorem~\ref{thm:gsc} also hold for thermal states.
Briefly, for any $\beta >0$, consider the thermal state
\begin{equation} \label{tstate}
\mathcal{P}_{L, \beta} = \frac{e^{- \beta H_L}}{{\rm Tr}[e^{- \beta H_L}]}
\end{equation}
For any bounded operator $A \in \mathcal{B}(\mathcal{H}_L)$,  denote by
$$
\langle A \rangle_{L, \beta} = {\rm Tr}[ A \mathcal{P}_{L, \beta}] 
$$
the expected value of $A$ in the thermal state $\mathcal{P}_{L, \beta}$. 
The following result is in \cite{NSS1}.

\begin{theorem} \label{thm:tsc} Suppose that the oscillator model $H_L$, as in (\ref{oscham}), is
random and satisfies Assumption A.
Then for any $\beta >0$, there exist $C < \infty$ and $\eta >0$ such that
\begin{equation} \label{stscd}
\mathbb{E} \left( \left| \langle W_x(z) W_y(z') \rangle_{L, \beta} - \langle W_x(z) \rangle_{L, \beta} \langle W_y(z') \rangle_{L, \beta} \right|^{1/2} \right)  \leq C|z|^{1/2} |z'|^{1/2} e^{- \eta |x-y|} 
\end{equation}
for all $L$, $x$, $y$, $z$, and $z'$.
\end{theorem}

Again, this static result holds in any dimension and at arbitrary disorder. The powers of 1/2 appearing above are 
necessary to temper stronger singularities which arise in the formulas for thermal state expectations.  In fact, the
bound (\ref{stscd}) follows from a slight modification of Lemma~\ref{lem:anymu}, see Proposition A.4(b) in \cite{NSS1}, which is specifically
adapted to these thermal expectations. Although not stated here, dynamic correlations
of thermal states can also be uniformly bounded, again in the large disorder regime. 
The result is analogous to (\ref{dgscd}) in Theorem~\ref{thm:gsc}, see \cite{NSS1} for details. 

As a final comment, it is further possible to prove that both static and dynamic ground state (and thermal state) correlations of position and momentum operators
decay as in Theorem~\ref{thm:gsc} and Theorem~\ref{thm:tsc} above, see \cite{NSS1} for details.

\subsection{Area Laws}  \label{sec:OSarealaws}

For a given random quantum model, another possible manifestation of many-body localization 
would be an area law for the bipartite entanglement entropy of the system's ground state.
Much like in the consideration of correlations from the previous subsection, it is clear that for product states 
the entanglement entropy vanishes. Verifying an area law then establishes that such a ground state has
relatively small entanglement, at least it is not extensive, and therefore the state more closely resembles that of
a non-interacting system.  By exploiting properties of the logarithmic negativity, an area law is proven for both
ground states and thermal states (to be interpreted properly) of random quantum oscillator models in \cite{NSS2}.

A more precise description of the various quantities involved follows. 
Fix a finite set $\Gamma \subset \mathbb{Z}^d$ and
take $L$ large enough so that $\Gamma \subset \Lambda_L$. For any such $L$,
form a bipartite decomposition of the Hilbert space $\mathcal{H}_L$ by writing 
$\mathcal{H}_L = \mathcal{H}_1 \otimes \mathcal{H}_2$ with
\begin{equation}
\mathcal{H}_1 = \bigotimes_{x \in \Gamma} L^2( \mathbb{R}) \quad \mbox{and} \quad \mathcal{H}_2 = \bigotimes_{x \in \Lambda_L \setminus \Gamma} L^2( \mathbb{R})
\end{equation}
Now, consider an oscillator Hamiltonian $H_L$, e.g.\ as in (\ref{oscham}), on this Hilbert space.
Under assumptions as before, denote by $\Omega_L$ its non-degenerate, normalized ground state. Furthermore, let
$\mathcal{P}_L = | \Omega_L \rangle \langle \Omega_L |$ denote the corresponding 
ground state projector, i.e., the orthogonal projection onto the subspace of
$\mathcal{H}_L$ spanned by $\Omega_L$. 

To define the entanglement entropy of the ground state with respect to this bipartite decomposition, we first trace
out the exterior degrees of freedom, i.e., we consider the non-negative, trace class operator on $\mathcal{H}_1$
given by
\begin{equation}
\mathcal{P}_L^1 = {\rm Tr}_{\mathcal{H}_2}[ \mathcal{P}_L]
\end{equation}
The main quantity of interest here is the entanglement entropy of this restriction, i.e., the von Neumann entropy of
$\mathcal{P}_L^1$ which we will denote by 
\begin{equation}
S( \mathcal{P}_L^1) = - {\rm Tr}[ \mathcal{P}_L^1 \ln( \mathcal{P}_L^1)] 
\end{equation}

In this context, an {\it area law} is a proof that $S(\mathcal{P}_L^1)$ grows like $| \partial \Gamma|$; the surface area of $\Gamma$.
Generally, it is believed that gapped quantum spin systems have ground states which satisfy an area law. This has been proven to be
the case in one dimension, see \cite{hast07}. More recently, it was shown that, again in one dimension, area laws for ground states follow from 
merely assuming exponential decay of correlations, see \cite{BrandaoHorodecki2012, Brand12}.  It is by now well-known that 
gapped quantum spin systems satisfy exponential decay of correlations in their ground state, see e.g. \cite{nachtergaele2005d, hastings2005}, and so the results in \cite{BrandaoHorodecki2012, Brand12} extend what was proven in \cite{hast07}. For $d>1$, however, there are no general results proving area laws for ground states under
either assumption. Since ground states that are minimally entangled can more easily be simulated by numerical methods, see e.g.\ \cite{Amico08, Ng07, Sk05}, 
it is therefore of great interest, e.g.\ to the quantum information community, to know when these area laws hold. 

For deterministic, gapped oscillator models, area laws have been proven, see \cite{CE,CEPD} and also the review \cite{ECP}.
In these works, the existence of the uniform gap is crucial. The following result is proven in \cite{NSS2}.

\begin{theorem} \label{thm:gsal} 
Suppose that the oscillator model $H_L$, as in (\ref{oscham}), is
random and satisfies Assumption A. Then
there exists $C' < \infty$ such that
$$
\mathbb{E} \left( S( \mathcal{P}_L^1)  \right)  \leq C' | \partial \Gamma| 
$$
for all $L$ with $\Gamma \subset \Lambda_L$.
\end{theorem}

It is important to stress a few points about this result. First, Theorem~\ref{thm:gsal} holds in any dimension and for arbitrary disorder.
Next, as discussed following Theorem~\ref{thm:gsc} in the previous subsection, there are classes of models
satisfying Assumption A for which the corresponding oscillator Hamiltonians $H_L$ are almost surely gapless in the thermodynamic limit.
In these cases, the fact that the corresponding ground states satisfy an area law is new, and
moreover, the mechanism which enables the proof is exactly the localization effects due to randomness.
Finally, as is described at the end of Section~\ref{sec:dl}, one can prove these area laws for more
general harmonic models if one assumes that the corresponding, effective single-particle Hamiltonian satisfies certain 
singular eigenfunction-correlator decay, see \cite{NSS2} for details.

The first crucial step in the proof of Theorem~\ref{thm:gsal} is to recognize that the bipartite entanglement entropy
associated to the ground state in these models, i.e.\ $S( \mathcal{P}_L^1)$, can be bounded from above by a quantity known as the {\it logarithmic negativity}
of the ground state. This important observation, 
which we state as Lemma~\ref{lem:vw} below, goes back to a work of Vidal and Werner in \cite{VW}.

Consider the following quantity. Let $\mathcal{P}$ be a non-negative, trace class operator on $\mathcal{H}_L$. The logarithmic negativity
of $\mathcal{P}$ with respect to the bipartite decomposition $\mathcal{H}_L = \mathcal{H}_1 \otimes \mathcal{H}_2$ is given by
\begin{equation}
\mathcal{N}( \mathcal{P}) = \log \left( \| \mathcal{P}^{T_1} \|_1 \right) 
\end{equation}
where $\mathcal{P}^{T_1}$ is the partial transpose of $\mathcal{P}$ and $\| \cdot \|_1$ is the corresponding 1-norm. In this setting, the
partial transpose is defined with respect to the natural conjugation map on $\mathcal{H}_1= L^2(\mathbb{R}^{\Gamma})$ and it is given by
\begin{equation}
(A \otimes B)^{T_1} = A^T \otimes B
\end{equation}
at least for product maps. Partial transposes for more general maps, as well as some of their basic properties,
are the topic of an appendix in \cite{NSS2}. The following result goes back to \cite{VW}. 

\begin{lemma} \label{lem:vw} If $\mathcal{P}$ is a rank-one projection on $\mathcal{H} = \mathcal{H}_1 \otimes \mathcal{H}_2$
and $\mathcal{P}^1 = {\rm Tr}_{\mathcal{H}_2}[ \mathcal{P}]$, then
$$
S( \mathcal{P}^1) \leq \mathcal{N}( \mathcal{P}) 
$$
\end{lemma}
Applying Lemma~\ref{lem:vw} to the ground state projectors, it is clear that to prove Theorem~\ref{thm:gsal} one need only estimate the
averaged logarithmic negativities. For oscillator models, this quantity is better suited for analysis. A review, which is beyond the scope
of this presentation, of some well-known calculations, specifically in the context of these infinite dimensional Hilbert spaces, is contained in \cite{NSS2}.
In fact, an immediate consequence of Lemma 4.1 from \cite{NSS2} is that 
there exists $C< \infty$ such that the following deterministic bound
\begin{equation} \label{lognegbd}
\mathcal{N}( \mathcal{P}_L) \leq C \sum_{x \in \Lambda_L} \sum_{y \in \Lambda_L \setminus \Gamma} \left| \langle \delta_x, h_L^{-1/2} \delta_y \rangle \right|
\end{equation} 
holds for any finite-volume oscillator model satisfying (\ref{constm+c}) and (\ref{randsc}), i.e.\ constant masses, constant couplings, and uniformly bounded
spring coefficients. As a result, Theorem~\ref{thm:gsal} now follows from Lemma~\ref{lem:anymu}.

More is true.  One can also define the logarithmic negativity of the thermal state $\mathcal{P}_{L, \beta}$ that was introduced in (\ref{tstate}). The following results, which can be interpreted as an {\it area law} at any positive
temperature, can be found in \cite{NSS2}.

\begin{theorem}\label{thm:tsal} Suppose that the oscillator model $H_L$, as in (\ref{oscham}), is
random and satisfies Assumption A. For any $\beta>0$, there exists $C' < \infty$ such that
\begin{equation} \label{tsal}
\mathbb{E} \left( \mathcal{N}( \mathcal{P}_{L, \beta}) \right) \leq C' | \partial \Gamma| 
\end{equation}
for all $L$ with $\Gamma \subset \Lambda_L$.
\end{theorem}

The proof of this result is again similar to the proof of Theorem~\ref{thm:gsal}. First, a deterministic analogue of
(\ref{lognegbd}) holds for these thermal states, but instead of $h_L^{-1/2}$ the bound involves the operators $\phi(h_L)$, where $\phi(t) = t^{-1/2} \tanh(\beta t^{1/2})$. This function is covered by Lemma~\ref{lem:anymu} and thus (\ref{tsal}) follows as in the proof of Theorem~\ref{thm:gsal}.

\end{document}